\documentclass[11pt]{article}
\usepackage{epsfig}

\begin{document}

\baselineskip 30pt

\begin{center}
\begin{LARGE}
{\bf Two Electron Quantum Dot - A Variational Treatment For The Ground State
                      }
\end{LARGE}
\end{center}

\vskip 0.2in

\begin{large}
\begin{center}
Subinoy Das  \\
Department of Physics, Indian Institute of Technology, Kanpur, U.P.--208016, India \\
email: subinoy@iitk.ac.in   \\ \mbox{} \\
Pallab Goswami \\
Department of Physics, Indian Institute of Technology, Kanpur, U.P.--208016, India \\
email: pallab@iitk.ac.in   \\ \mbox{} \\
and \\
J.K. Bhattacharjee\footnote{Author to whom all correspondence should be addressed}  \\
Department of Theoretical Physics, Indian Association for Cultivation of Science \\
Calcutta -- 700032, India \\
email: tpjkb@mahendra.iacs.res.in  \\ \mbox{} \\
{Dated:14th January, 2002} \\ \mbox{} \\

\end{center}
\end{large}
\baselineskip 18pt

\parindent 0.3in

A variational treatment for a two-electron quantum dot (the artificial helium atom) is
proposed which leads to exact answer for the ground state energy. Depending on the magnetic
field value the singlet-triplet and triplet-triplet transitions of the ground state take
place, which are captured in our solution.Using the same variational technique we find
corrections to wave function and energy and the transition field srengths in a realistic
dot where electron wave function has a finite extent in the direction perpendicular to the
x-y plane in which usually 2-D dot electrons are confined. Using the variational wave
function we show that photoemission cross-section as a function of magnetic field has sharp
discontinuities, which can be used for experimental verification of the singlet-triplet
transitions.

Quantum dots [1] are little two-dimensional islands of electrons, which are laterally
confined by an artificial potential. They can be thought of as artificial atoms with the
field of nucleus replaced by an imposed external potential.The artificial hydogen atom
is a single electron in a two dimensional circular geometry confined by a harmonic potential.
The problem becomes interesting in the presence of magnetic field in the perpendicular
direction and wave funcions for this case were worked out by Fock [2] shortly after the
Schroendiger equation was established. The artificial 'helium atom' problem was, however
taken up more than forty years later. In an extensive numerical work Maksym and Chakraborty
[3] and Wagner et al [4] found extremely interesting effect of the competition between
magnetic field and Coulomb repulsion between electrons in two-electron quantum dots. In
particular these authors found that the ground state can change character as the magnetic
field changes, leading to singlet-triplet transitions. Six years later Dineykhan and Nazmitdinov
[5] using the formidable tools of constructing equivalent hamiltonians in oscillator representations
solved the problem for a two-electron quantum dot exactly. In this communication we note
the fact that in the real helium atom problem, excellent results for the ground state are
obtained by using a variational principle and exploit that to set up a variational calculation
for this artificial helium atom. The ground state energy turns out to be the same as obtained
from the exact treatment. Employing the same variational techniques we find the required correction
for energy, wave function and the magnetic field strengths at which transitions are taking
place for a more realistic quantum dot in which elctron wave function has finite extent
in the third direction. This is done by imposing a more stronger harmonic confinement in
the z-direction. We also show that the photoemission cross-section calculated using the
variational wave function has sharp discontinuities at the magnetic field values where
the transitions occur which has been proposed to probe it experimentally [6].

For two electrons confined by a parabolic potential in two-dimensional plane,the hamiltonian
is (gauge $A_{x}=\frac{1}{2}By$, $A_{y}=-\frac{1}{2}Bx$)
\begin{equation}
H  =  \sum_{j=1}^{2} [-\frac{\hbar^{2}}{2m^{\ast}}\nabla_{j}^{2} +\frac{\omega_{c}}{2}(-i\hbar\nabla_{\phi_{j}})+\frac{1}{2}m^{\ast}\Omega^{2}\rho_{j}^{2}] +\frac{e^{2}}{4\pi\epsilon\epsilon_{0}}\frac{1}{\mid \vec{\rho_{1}} -\vec{\rho_{2}}\mid}
\end{equation}
and if the confinement in the z-direction is considered the hamiltonian is modified by the term
\begin{equation}
H_{z} = -\frac{\hbar^{2}}{2m^{\ast}}\frac{d^{2}}{dz^{2}}+\frac{1}{2}m^{\ast}\omega_{z}^{2}z^{2}
\end{equation}
and the Coulomb term becomes
$\frac{e^{2}}{4\pi\epsilon\epsilon_{0}}\frac{1}{\mid \vec{r_{1}} -\vec{r_{2}}\mid}$.
In the $H$ above $\vec{\rho_{1}}$ and $\vec{\rho_{2}}$ are the two dimensional position vectors of the
two electrons ($\vec{r_{1}}$ and $\vec{r_{2}}$ are three dimensional position vectors),
$\omega_{c}=\frac{eB}{m^{\ast}}$ is the cyclotron frequency, $\omega_{0}$ is the frequency
of the confining potential in x-y plane ($\omega_{z}$ is the confining frequency in the
z-direction and $\omega_{z}\gg\omega_{0}$), $m^{\ast}$ is the effective mass of the electron
in the semiconductor, $\epsilon$ is the dielectric constant and $\tilde{g}$ is effective
Lande factor for the semiconductor and
$\Omega^{2} = (\omega_{0}^{2} +\frac{\omega_{c}^2}{4})$. We transform to the center of
mass coordinate $\vec{\rho_{c}}=\frac{1}{2}(\vec{\rho_{1}}+\vec{\rho_{2}})$ and the
relative coordinate $\vec{\rho_{rel}}=(\vec{\rho_{1}}-\vec{\rho_{2}})$ and write the
hamiltonian in equation (1) as $ H=H_{c}+H_{rel}$, where $H_{rel}=H_{rel}^{0}+H_{int}$
and
\begin{eqnarray}
H_{c} & = & -\frac{\hbar^{2}}{4m^{\ast}}\nabla_{c}^{2}+
\frac{\omega_{c}}{2}(-i\hbar\nabla_{\phi_{c}})+m^{\ast}\Omega^{2}\rho_{c}^{2} \\
H_{rel}^{0} & = & -\frac{\hbar^{2}}{m^{\ast}}\nabla_{rel}^{2}+\frac{\omega_{c}}{2}(-i\hbar\nabla_{\phi_{rel}})+\frac{1}{4}m^{\ast}\Omega^{2}\rho_{rel}^{2} \\
H_{int} & = & \frac{e^{2}}{4\pi\epsilon\epsilon_{0}}\frac{1}{\rho_{rel}}
\end{eqnarray}
The wave function $\Psi(\vec{\rho_{c}},\vec{\rho_{rel}})$ will clearly separate as
$\psi_{1}(\vec{\rho_{c}})\psi_{2}(\vec{\rho_{rel}})$ with the energy eigenvalue E splitting
as $E=E_{c} +E_{rel}$ where
\begin{eqnarray}
H_{c}\psi_{1}(\vec{\rho_{c}}) & = & E_{c}\psi_{1}(\vec{\rho_{c}})  \\
H_{rel}\psi_{2}(\vec{\rho_{rel}}) & = & E_{rel}\psi_{2}(\vec{\rho_{rel}})\end{eqnarray}
$H_{c}$ and $H_{rel}^{0}$ are hamiltonians of single electron quantum dots with the masses of the
electrons given by 2m* and m*/2 respectively. Consequently the exact answers are known for these
parts and we have
\begin{eqnarray}
E_{c} & = & (2N+\mid M \mid+1)\hbar\Omega-\frac{\mid M \mid\hbar\omega_{c}}{2} \nonumber \\
E_{rel}^{0} & = & (2n+\mid m \mid +1)\hbar\Omega-\frac{\mid m \mid\hbar\omega_{c}}{2}
\end{eqnarray}
with the wave functions given by
\begin{eqnarray}
\psi_{1NM}(\vec{\rho_{c}}) & = & \sqrt {\frac{\Gamma(N+1)}{2^{\mid M \mid+1}\pi \tilde{a_{H}^{2}}\Gamma(N+\mid M \mid+1)}}(\frac{\rho_{c}}{\tilde{a_{H}}})^{\mid M \mid}L_{N}^{\mid m \mid}(\frac{\rho_{c}^{2}}{2\tilde{a_{H}^{2}}})\times
exp[-\frac{\rho_{c}^{2}}{4\tilde{a_{H}^{2}}}-iM\phi_{c}]
\end{eqnarray}
\begin{eqnarray}\psi_{2nm}(\vec{\rho_{rel}}) & = & \sqrt {\frac{\Gamma(n+1)}{2^{\mid m \mid+1}\pi a_{H}^{2}\Gamma(n+\mid m \mid+1)}}
(\frac{\rho_{rel}}{a_{H}})^{\mid m \mid}L_{n}^{\mid m \mid}(\frac{\rho_{rel}^{2}}{2a_{H}^{2}})\times \nonumber \\
&   &  exp[-\frac{\rho_{rel}^{2}}{4a_{H}^{2}}-im\phi_{rel}]\end{eqnarray}
Here radial quantum numbers N, n are positive integers and angular momentum quantum
numbers M, m can take all possible integral values. Length scales are set by
$\tilde{a_{H}}$ and $a_{H}$ and $\tilde{a_{H}^{2}}=\frac{\hbar}{4m^{\ast}\Omega}=
\frac{a_{H}^{2}}{4}$. In addition to these if we consider the Zeeman term then
there would be an energy contribution
\begin{equation}
E_{spin}=\tilde{g}\mu_{B}B\sum_{j=1}^{2} S^{z}_{j}=\tilde{g}\mu_{B}BS^{z}_{total}\end{equation}
where $S^{z}_{j}$ is the z-component of the spin operator of the j-th electron and $\tilde{g}$
is the effective Lande factor for the semiconductor. When permutation of electrons take place
$\vec{\rho_{rel}}\rightarrow-\vec{\rho_{rel}}$ and antisymmetry requirement implies for
odd m values triplet and for even m values singlet states and
$E_{spin}=\tilde{g}\mu_{B}B(1-(-1)^{m})$.

However there is no exact answer for $H_{int}$ and consequently approximation
techniques have to be resorted to. Our observation is that the variational
principle which one employs to calculate the ground state of helium atom
should be effective here as well. This is what prompts our trial wave
function. The center of mass motion does not enter the picture, the ground
state of that being fixed by $N = M = 0$. In the absence of the Coulomb
interaction, the ground state of the relative motion would be given by $n = m = 0$.
It is important to note that in the absence of confinement (i.e.$\omega_{0}=0$),
the energy levels are independent of m, but in the presence of confining potentials,
for a given value of n, there is an interesting dependence on the azimuthal quantum
number m. Magnetic field tries to compress the wave function i.e., to decrease the
separation of the electrons where as to minimize the Coulomb repulsion electrons
would tend to increase their separation. Optimization of these competing effects
i.e., minimization of the total energy takes place at different m values depending on
the strength of the magnetic field. Now to maintain the anti-symmetry of the total wave
function the spin state changes between triplet to singlet corresponding to change in
m from odd to even value and vice versa as evident from the $E_{spin}$ expression. At
strong magnetic field due to spin polarization, spin states would be triplets; thus
ground state ceases to show further singlet- triplet transition and only triplet-triplet
transitions take place. Consequently, in writing down the ground state wave function,
we respect this fact by anticipating an n = 0 form for the radial wave function but allowing
for an azimuthal dependence as $\exp(-im\phi_{rel})$. As the dominating part of
the potential terms in $H^{0}_{rel}+H_{int}$ is the parabolic one, the effect of
the Coulomb part can be accounted by introducing the length scale as the variational parameter
in analogy with the helium atom problem.  This inspires the variational trial wave function
(normalized)\begin{equation}
\psi_{0m}(\vec{\rho_{rel}}) = \sqrt {\frac{1}{2^{\mid m \mid+1}\pi \beta^{2}\Gamma(\mid m \mid+1)}}(\frac{\rho_{rel}}{\beta})^{\mid m \mid}exp[-\frac{\rho_{rel}^{2}}{4\beta^{2}}-im\phi_{rel}]\end{equation}
With this trial function, there will be four contributions to $E_{rel}$, which we write as
\begin{eqnarray}
E_{1}(\beta,m) & = & <\psi_{0m}\mid - \frac{\hbar^{2}}{m^{\ast}}\nabla_{rel}^{2}\mid\psi_{0m}>  =  \frac{\hbar^{2}}{2m^{\ast}\beta^{2}}(\mid m \mid+1) \nonumber \\
E_{2}(\beta,m) & = & <\psi_{0m}\mid\frac{\omega_{c}}{2}(-i\hbar\nabla_{\phi_{rel}})\mid\psi_{0m}>  =  -\frac{\omega_{c}m\hbar}{2} \nonumber \\
E_{3}(\beta,m) & = & <\psi_{0m}\mid\frac{m^{\ast}\Omega^{2}\rho^{2}_{rel}}{4}\mid\psi_{0m}>  = \frac{m^{\ast}\Omega^{2}\beta^{2}}{2}(\mid m \mid+1) \nonumber \\
E_{4}(\beta,m) & = & <\psi_{0m}\mid\frac{e^{2}}{4\pi\epsilon_{0}\epsilon\rho_{rel}}\mid\psi_{0m}>  =  \frac{e^{2}\Gamma(\mid m \mid+\frac{1}{2})}{4\pi\epsilon_{0}\epsilon\beta\sqrt{2}\Gamma(\mid m \mid+1)}\end{eqnarray}
This leads to \begin{equation}
E_{rel}(\beta,m)=(\mid m \mid +1)[\frac{\hbar^{2}}{2m^{\ast}\beta^{2}}+\frac{m^{\ast}\Omega^{2}\beta^{2}}{2} ] -\frac{\omega_{c}m\hbar}{2}+\frac{e^{2}\Gamma(\mid m \mid+\frac{1}{2})}{4\pi\epsilon_{0}\epsilon\beta\sqrt{2}\Gamma(\mid m \mid+1)}\end{equation}
Minimization of this energy with respect to $\beta$ leads to
\begin{equation}x^{4}-\frac{a_{H}}{a^{\ast}}\frac{1}{\sqrt{2}}\frac{\Gamma(\mid m \mid +\frac{1}{2})}{\Gamma(\mid m \mid +2)}x -1 = 0\end{equation}
where $x=\frac{\beta}{a_{H}}$ and $a^{\ast}=\frac{4\pi\epsilon\epsilon_{0}\hbar^{2}}{m^{\ast}e^{2}}$.
We solved equation (15) numerically for GaAs bulk conduction band electron parameters
($g=-0.44$, $\frac{m^{\ast}}{m_{e}}=0.067$, $\epsilon\approx{13}$) and the results are
shown in fig.1 and fig.2. We have taken the typical confinement energy to be 4meV and
varied B from 0 to 12T. In fig.1 $\frac{E_{rel}+E_{spin}}{\hbar\omega_{0}}$ vs. B is plotted.
The transitions $m=0\rightarrow m=1$, $m=1\rightarrow m=2$ occur respectively at magnetic
field strengths 1.3T, 6.1T (approximately). The mean square separation of two elctrons is
given by $<\psi_{0m}\mid \rho_{rel}^{2}\mid\psi_{0m}>=2(\mid m \mid +1)\beta^{2}$ and root
mean square separation $d=\frac{\sqrt{<\psi_{0m}\mid \rho_{rel}^{2}\mid\psi_{0m}>}}{a^{\ast}}$
vs. B is plotted for different m values in fig.2 showing the discontinuities brought about in
the inter-electron separation by the transitions.

However for a more realistic quantum dot the finite extent of the electronic wave function in the
third direction has to be taken into account and we achieve this by imposing another harmonic
confinement in the z-direction and employ our variational tools to get the characteristic equation.
With the $H_{z}$ part we have the wavefunction modified by
\begin{equation}\psi_{3}(z)=\frac{1}{\sqrt{2^{n}\Gamma(n_{z}+1)\pi^{1/2}\lambda}}\exp(-\frac{z^{2}}{2\lambda^{2}})H_{n}(\frac{z}{\lambda})\end{equation}
and the energy is modified by the term $E_{z}=(n_{z}+\frac{1}{2})\hbar\omega_{z}$. Separating the
hamiltonian again in center of mass and relative coordinates and following previous arguments it
can be clearly seen that the ground state would have quantum numbers $N_{z}^{c}=0$ and $n_{z}^{rel}=0$.
We introduce the variational parameters $\beta_{1}$ and $\beta_{2}$ respectively replacing $a_{H}$ and
$\lambda_{rel}$ and after energy minimization get coupled characteristic equations for $x=\frac{\beta_{1}}
{a_{H}}$ and $y=\frac{\beta_{2}}{\lambda_{rel}}$ where $\lambda_{rel}^{2}=2\lambda^{2}=\frac{2\hbar}{m^{\ast}\omega_{z}}$.
The characteristic equations are
\begin{equation}x^{4}[1-\frac{2}{y^{3}}(\frac{a_{H}}{a^{\ast}})(\frac{a_{H}}{\lambda_{rel}})^{3}\frac{\Gamma(\mid m \mid+1)}{\Gamma(\frac{5}{2}+\mid m \mid)}2F_{1}(\frac{3}{2},2+\mid m \mid,\frac{5}{2}+\mid m \mid,1-2(\frac{x}{y})^{2}(\frac{a_{H}}{\lambda_{rel}})^{2})]=1\end{equation}
\begin{eqnarray}y^{4}-y(\frac{\lambda_{rel}}{a^{\ast}})\frac{\Gamma(\mid m \mid+1)}{\Gamma(\frac{3}{2}+\mid m \mid)}2F_{1}(\frac{1}{2},1+\mid m \mid,\frac{3}{2}+\mid m \mid,1-2(\frac{x}{y})^{2}(\frac{a_{H}}{\lambda_{rel}})^{2})+2\frac{x^{2}}{y}\frac{a_{H}^{2}}{a^{\ast}\lambda_{rel}}\frac{\Gamma(\mid m \mid+2)}{\Gamma(\frac{5}{2}+\mid m \mid)}\times \nonumber \\
2F_{1}(\frac{3}{2},2+\mid m \mid,\frac{5}{2}+\mid m \mid,1-2(\frac{x}{y})^{2}(\frac{a_{H}}{\lambda_{rel}})^{2})=1\end{eqnarray}
We have solved these coupled equations simultaneously with GaAs parameters and found that singlet-triplet
transitions are taking place at higher magnetic field strengths and these results are shown in Fig.3.
These results can be understood from the physical ground that with the introduction of
harmonic confinement in the z-direction x values decrease significantly compared to the
exact 2-D situation (Fig.4) leading to increase in Coulomb energy and to minimize the total energy
the required transitions need higher energy contribution from the magnetic field. Therefore,
the transitions occur at higher fields strengths. All these results are in good agreement with
the exact solution of Dineykhan and Nazmitdinov [5].

Usually the energy levels in the semiconductor quantum dots are probed by the FIR spectroscopy.
Due to the long wavelength of the electromagnetic field (compared to the dot size), there is no
appreciable change in the electric field across the dot and hence, the elctric field couples only
to $H_{c}$ through the contribution
$e\vec{E}\cdot({\vec{\rho}}_{1}+{\vec{\rho}}_{2})\exp(i\omega t)=2e\vec{E}\cdot\vec{\rho_{c}}\exp(i\omega t)$,
where $\vec{E}$ is a constant electric field (dipole approximation).
As $H_{rel}$ is not coupled to the field, FIR spectroscopy does not probe the effects of
inter-electron repulsion. However if we consider the photoemission from the dot to the vacuum
by irradiating the dot with high energy photons, then $H_{rel}$ is coupled to the external
fields (as the wave length is comparable to the dot size, dipole approximation is not valid)
and the singlet-triplet and triplet-triplet transitions are manifested as sharp discontinuities
in the photoemission cross-section. The calculation of the photoemission cross-section involves
the evaluation of a matrix element corresponding to the electromagnetic field operator causing
the emission from the initial state to final state and thus depends on the details of the initial
and final state wave functions. This dependence on the initial and final state introduces the m
dependence of  the cross-section and in this case we do not replace $\exp(i\vec{k}\cdot\vec{r})$
by unity as done in the dipole approximation. The operator is treated exactly and the final form
of the differntial cross-section calculated using our analytically found trial wave function becomes
\begin{equation}\frac{\frac{d\sigma}{d\phi_{\vec{q}}}}{\sigma_{0}} = \frac{2^{8+3m}}{\Gamma(\mid m \mid +1)}[1-(\mid m \mid +1)\frac{\Omega}{\omega}+\frac{m}{2}\frac{\omega_{c}}{\omega}]
\sin^{2}\theta\cos^{2}\phi\frac{(x^{2}+2)^{2\mid m \mid}(Ka_{H})^{2\mid m \mid}}{x^{2\mid m \mid-2}(x^{2}+6)^{2\mid m \mid+2}}\exp[-2\frac{K^{2}a_{H}^{2}(3x^{2}+2)}{(6+x^{2})}]\end{equation}
Here, $\vec{q}$ is the wave vector of the emitted electron, $\vec{k}$ is the wave vector of the incident
photon, $\hbar\vec{K}=\hbar(\vec{k}-\vec{q})$ is the momentum transferred and
$\theta$ and $\phi$ are respectively the angles $\vec{q}$ makes with $\vec{k}$
and $\vec{k}\hat{n}$ plane where $\hat{n}$ is the unit polarization vector of
the incident photon. $\sigma_{0}=\frac{e^{2}}{c{a^{\ast}}^{2}}(\frac{m_{e}}{m^{\ast}})^{2}$ is a
constant extracted to express the differntial cross-section expression in a dimensionless form.
So, \begin{equation}K^{2}=k^{2}+q^{2}-2kq\cos\theta \end{equation}
holds and $\cos\phi=\sin\theta_{\hat{n}}\cos(\phi_{\hat{n}}-\phi_{\vec{q}})$
and $\cos\theta=\sin\theta_{\vec{k}}\cos(\phi_{\vec{k}}-\phi_{\vec{q}})$. Therefore, it is very clear
from the expression for the differential cross-section that it depends significantly
on the direction of incidence and polarization. This for some simple cases can be
illustrated easily. If $\vec{k}$ is parallel to z-axis then $\cos\theta=0$ and
$\cos\phi=\cos\phi_{\vec{q}}$ or $\cos\phi=\sin\phi_{\vec{q}}$ as $\hat{n}$ is parallel to x or
y-axis. So, the angular distribution is proportional to $\cos^{2}\phi_{\vec{q}}$ or $\sin^{2}\phi_{\vec{q}}$ and if
it is the case of circular polarization then the angular distribution is proportional to $(\cos^{2}\phi_{\hat{n}}\cos^{2}\phi_{\vec{q}}+\sin^{2}\phi_{\hat{n}}\sin^{2}\phi_{\vec{q}})$ and
only if $\hat{n}=\frac{(\hat{x}\pm \hat{y})}{\sqrt{2}}$ then it becomes isotropic.
But, when the $\vec{k}$ lies in the x-y plane then with all the cases of circular polarization we shall
have angular dependence. It also becomes apparent from the expression that emission count is
larger in the direction of polarization compared to other cases and if the photon is
linearly polarized in the z-direction then there is no emission. So, depending on
the 'm' values of the ground state as a function of magnetic field the cross-section
would have different angular distribution as well as discontinuities characterizing
transitions of the ground state. Based on these one can probe now these transitions
experimentally. In the context of experiment the frequency of the photon has to be carefully chosen as
well as those directions mentioned above and details of these shall be provided
elsewhere [6] where we also show the modifications in the context of more realistic
quantum dots due to finite thickness. For illustration we plot the above expression for
the transition $m=0\rightarrow m=1$ and $m=1\rightarrow m=2$ in fig.5.

It should be noted that, using the proper variational wave function for the ground state we have
reproduced the exact solution of Dineykhan and Nazmitdinov [5]. The essential features of
the interacting two electron ground state of the semiconductor quantum dots, i.e.,
the singlet-triplet transitions with the change in the magnetic field strength and
triplet-triplet transitions at very strong magnetic field are explained using our result.
In the triplet state the angular momentum of the ground state is m = 2p + 1,where p is a positive
integer and this is weak version of the magic number concept. As, in our choice of the trial wave
function the symmetry under permutation is properly taken into account and the orthonormalization
is built in, proceeding in the same way it is quite trivial to find the higher excited states with
nonzero radial excitations i.e., $n\neq0$ . The competing effects of the magnetic field and Coulomb
repulsion are also demonstrated in terms of the variational length scale, which captures the
discontinuous changes in the inter-electron separation at transition points. Also, by same kind of
variational calculation the wave function and energy found for a dot with finite thickness
are in good agreement with the physically anticipated results. From the calculated photoemission
cross section using the above mentioned trial wave function we have also analytically demonstrated
the sharp discontinuities and the difference in the angular distribution at transition points which
can be used to probe these transitions for the parabolic dot in addition to the found discontinuities
from the magnetization measurement.

\bibliographystyle{plain}

\begin{figure}
\begin{center}
\epsfig{file=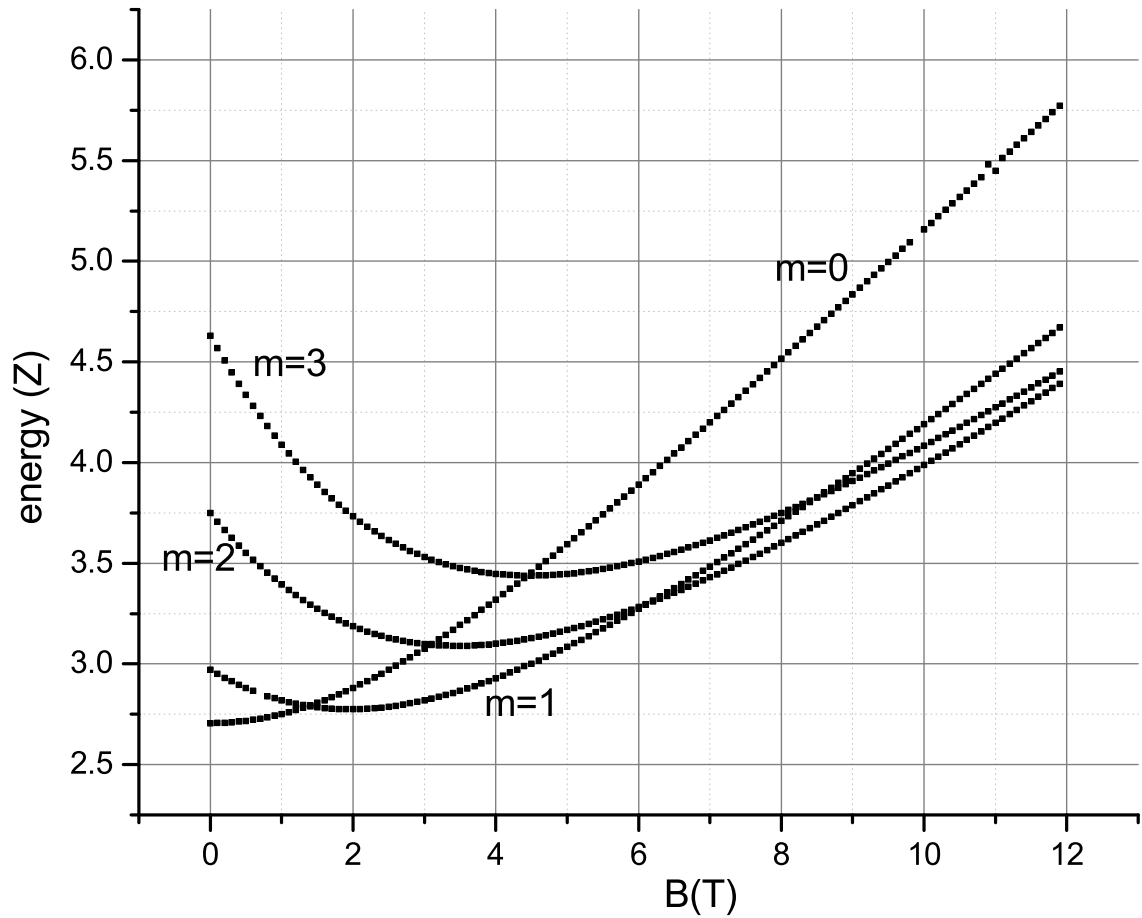,width=1.0\linewidth}
\end{center}
\vskip 0.5in
\caption{$Z=\frac{E_{rel}+E_{spin}}{\hbar\omega_{0}}$ vs. B(T) is plotted for different 'm' values and $m=0\rightarrow m=1$, $m=1\rightarrow m=2$ transitions are taking place at $B=1.3T$ and $B=6.1T$ respectively and also other energy level crossings are present.}
\label{fig:fig1}
\end{figure}

\begin{figure}
\begin{center}
\epsfig{file=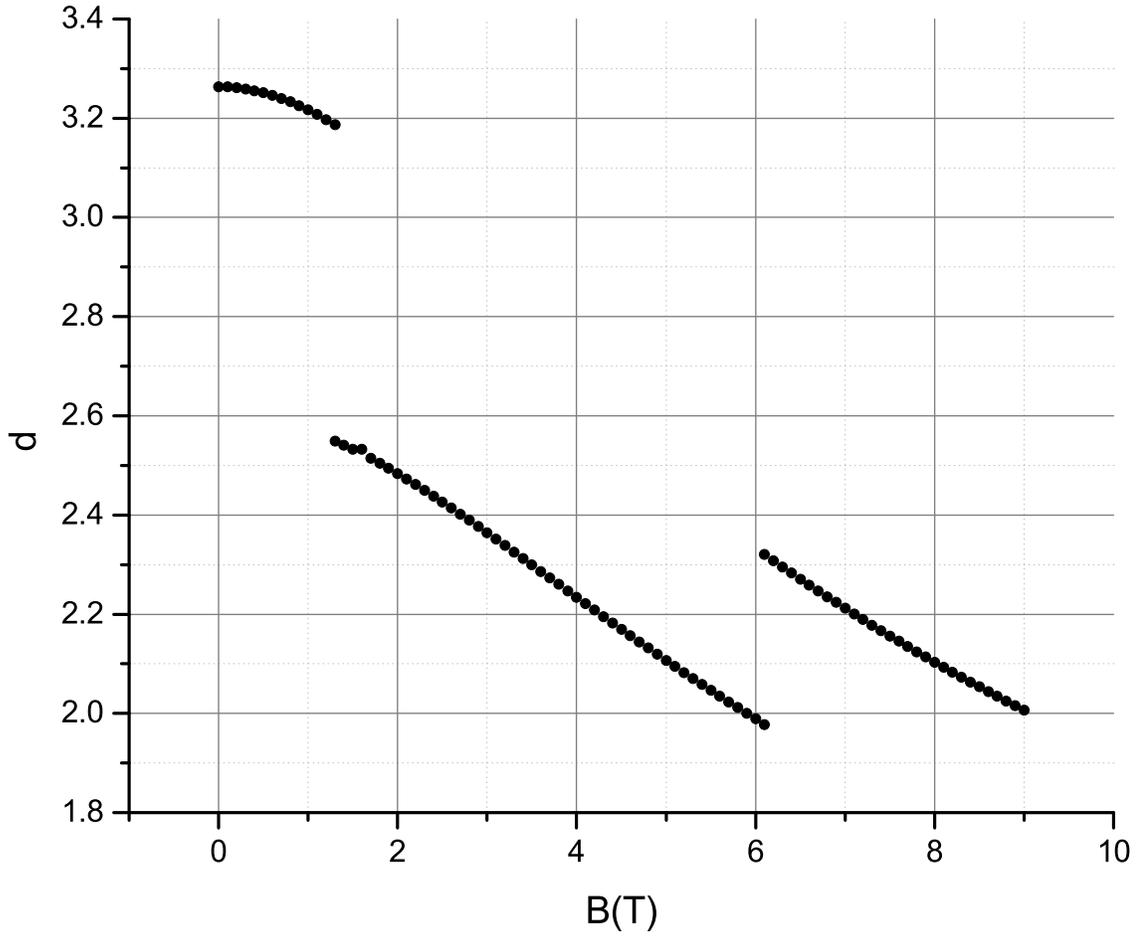,width=1.0\linewidth}
\end{center}
\vskip 0.5in
\caption{$d=\frac{\sqrt{<\psi_{0m}\mid \rho_{rel}^{2}\mid\psi_{0m}>}}{a^{\ast}}$ vs. B(T) is plotted for differnt 'm' values of the ground state and the transitions are captured in the discontinuous changes of d}
\label{fig:fig2}
\end{figure}

\begin{figure}
\begin{center}
\epsfig{file=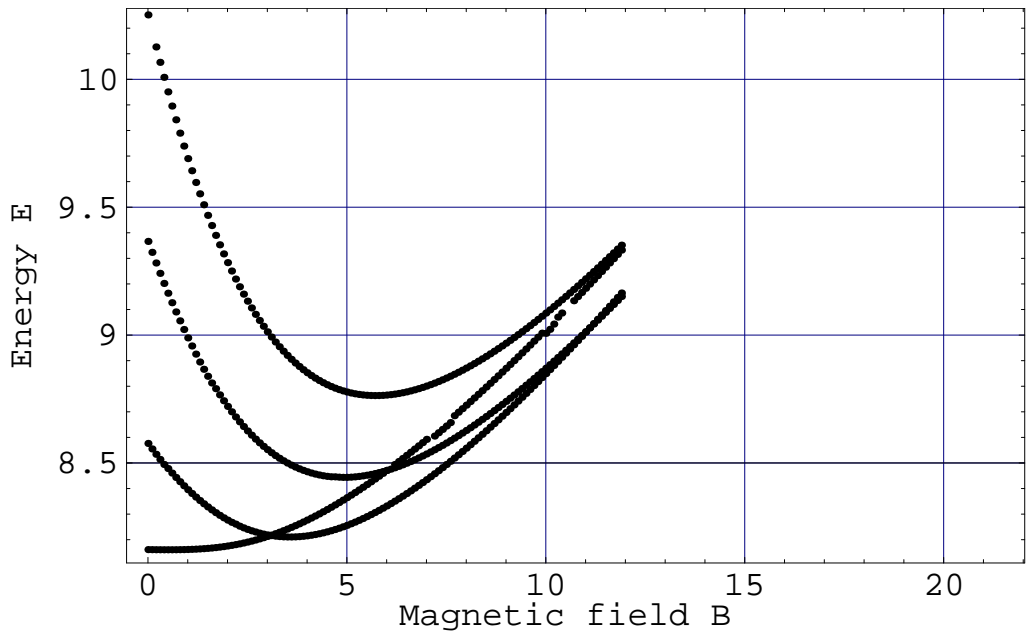,width=1.0\linewidth}
\end{center}
\vskip 0.5in
\caption{$E=\frac{E_{rel}+E_{spin}}{\hbar\omega_{0}}$ vs. B(T) plotted for different 'm' values when finite thickness of the dot is taken into account.For this purpose $\frac{\omega_{z}}{\omega_{0}}=9$ has been taken.}
\label{fig:fig3}
\end{figure}

\begin{figure}
\begin{center}
\epsfig{file=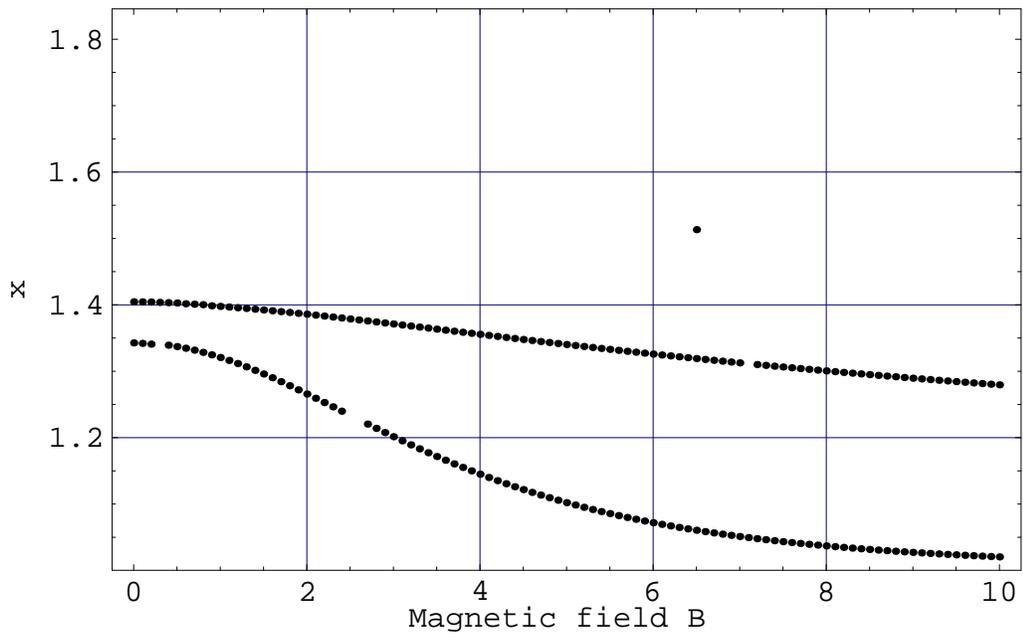,width=1.0\linewidth}
\end{center}
\vskip 0.5in
\caption{$x$ vs. B(T) is plotted for $m=0$ for both the 2-D and realistic 3-D dot and it is clearly found that throughout the range of magnetic field x values have decreased for the later one. }
\label{fig:fig4}
\end{figure}

\begin{figure}
\begin{center}
\epsfig{file=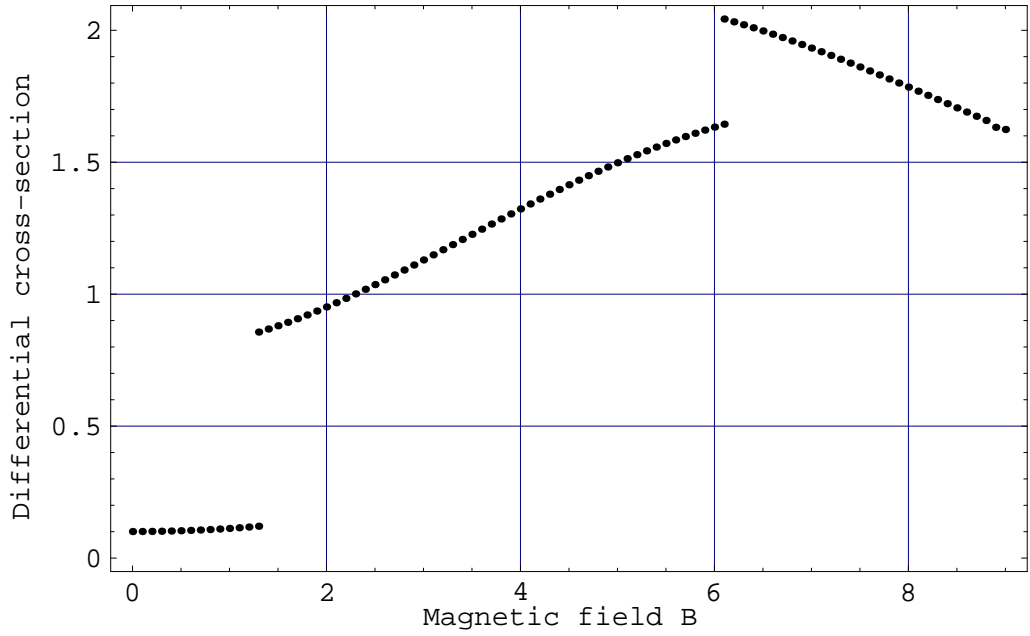,width=1.0\linewidth}
\end{center}
\vskip 0.5in
\caption{$\frac{\frac{d\sigma}{d\phi_{\vec{q}}}}{\sigma_{0}}$ vs. B(T) is plotted for different 'm' values of the ground state as the B is varied and showing discontinuities as characteristic of the transitions. From the plot it is found that percentage change for $m=1 \rightarrow m=2$ is $\approx 4.5$ times smaller compared to $m=0 \rightarrow m=1$ transition and also it is evident from the plot that at a fixed frequency behaviors of different 'm' cross-sections are going to change and for this reason both $\omega$ and B have to be varied to observe all the transitions properly.}
\label{fig:fig5}
\end{figure}

\end{document}